# Near-Unity Molecular Doping Efficiency in Monolayer MoS$_2$

Milad Yarali[1,2,+], Yiren Zhong[2,3,+], Serrae N. Reed[1,2], Juefan Wang[4], Kanchan A. Ulman[5], David J. Charboneau[3], Julia B. Curley[3], David J. Hynek[1,2], Joshua V. Pondick[1,2], Sajad Yazdani[1,2], Nilay Hazari[3], Su Ying Quek[4,5], Hailiang Wang[2,3], Judy J. Cha[1,2]

1. Department of Mechanical Engineering and Materials Science, Yale University, New Haven, CT 06511, USA
2. Energy Sciences Institute, Yale West Campus, West Haven, CT 06516, USA
3. Department of Chemistry, Yale University, New Haven, CT 06511, USA
4. Department of Physics, National University of Singapore, 117551, Singapore
5. Centre for Advanced 2D Materials, National University of Singapore, Block S14, Level 6, 6 Science Drive 2, 117546, Singapore

[+] These authors contributed equally to this work.

**Surface functionalization with organic electron donors (OEDs) is an effective doping strategy for two-dimensional (2D) materials, which can achieve doping levels beyond those possible with conventional electric field gating. While the effectiveness of surface functionalization has been demonstrated in many 2D systems, the doping efficiencies of OEDs have largely been unmeasured, which is in stark contrast to their precision syntheses and tailored redox potentials. Here, using monolayer MoS$_2$ as a model system and an organic reductant based on 4,4'-bipyridine (DMAP-OED) as a strong organic dopant, we establish that the doping efficiency of DMAP-OED to MoS$_2$ is in the range of 0.63 to 1.26 electrons per molecule. We also achieve the highest doping level to date in monolayer MoS$_2$ by surface functionalization and demonstrate that DMAP-OED is a stronger dopant than benzyl viologen, which was the previous best OED dopant. The measured range of the doping efficiency is in good agreement with the values predicted from first-principles calculations. Our work provides a basis for the rational design of OEDs for high-level doping of 2D materials.**

Controlling carrier densities in semiconductors is essential for producing functional devices and can be achieved using various doping techniques. However, conventional doping methods, such as substitutional doping or ion implantation, do not work well for two-dimensional (2D) materials that are being explored as alternative platforms for the next-generation of logic and memory devices[1,2]. Instead, owing to the high surface area, surface functionalization with organic electron donors (OEDs) is an effective approach to tune carrier density in 2D materials. This strategy has been successfully demonstrated in carbon nanotubes, graphene, and transition metal dichalcogenides for both electron and hole doping to generate materials with a wide range of doping levels[3-21]. For $MoS_2$ in particular[22], the two-electron reduced form of benzyl viologen ($BV^0$) can dope $MoS_2$ degenerately with an electron sheet density of $\sim 1.2 \times 10^{13}$ cm$^{-2}$.

Despite the remarkable success of surface functionalization, a significant mismatch persists between the precision syntheses of OEDs with tailored structures and properties and our understanding of their doping efficiencies and effects on 2D materials. Accurate measurements of their doping powers are largely lacking, without which the full potential of OEDs as dopants cannot be realized through rational design. Here, we use monolayer $MoS_2$ flakes as a model system and measure the doping power of an organic dopant based on 4,4'-bis(dimethylamino)bipyridine (DMAP-OED)[23], which functionalizes the surface of $MoS_2$. The change in the carrier density after doping was measured by comparing the field-effect transistor (FET) characteristics of $MoS_2$ before and after functionalization, while the number of DMAP-OED molecules present on $MoS_2$ was quantified by X-ray photoelectron spectroscopy (XPS) and atomic force microscopy (AFM). We report a doping efficiency ranging between 0.63 and 1.26 electrons per molecule for DMAP-OED to $MoS_2$ and achieve a degenerate doping level with a carrier density of $5.8 \pm 1.9 \times 10^{13}$ cm$^{-2}$ at the maximum functionalization conditions. This is four times greater than the best current system based on $BV^0$ [22]. The doping levels achieved with DMAP-OED are well beyond those possible by field-effect gating and offer opportunities to access a wide range of electronic phases that are tuned by the electron density[24].

DMAP-OED is a neutral reductant which is stable and soluble in organic solvents under a nitrogen or argon atmosphere[23]. Figure 1a shows the molecular structures and cyclic voltammograms (CVs) of DMAP-OED and $BV^0$. The CV of DMAP-OED shows a single, reversible two-electron wave at -1.22 V *vs.* the saturated calomel electrode (SCE). This reduction potential is significantly more negative than that of $BV^0$, which undergoes two discrete one-electron events at -0.33 V and -0.72 V *vs.* SCE. Based on their redox potentials, we hypothesized that DMAP-OED would be a stronger dopant for $MoS_2$ than $BV^0$. In agreement with the CV data, density functional theory (DFT) calculations indicate that a single DMAP-OED molecule adsorbed on a freestanding, monolayer $MoS_2$ (Fig. 1b) should transfer 0.99 electrons to $MoS_2$, while a $BV^0$ molecule should transfer 0.69 electrons to $MoS_2$ in the same adsorption configuration (Supplementary Fig. S1; Methods).

Triangular flakes of monolayer $MoS_2$ were synthesized on $Si/SiO_2$ substrates by chemical vapor deposition (Supplementary Fig. S2 and Methods)[25]. As-grown $MoS_2$ flakes were transferred to a fresh $SiO_2$ (285 nm) / Si ($p^+$) substrate and back-gated $MoS_2$ FETs were fabricated using standard e-beam lithography and e-beam evaporation of Ti/Au contacts (fabrication details in Methods). For molecular doping, a 10 mM solution of DMAP-OED in acetonitrile was drop-cast on the $MoS_2$ FETs for 10 minutes inside an Ar-filled glove box. The devices were then rinsed with acetonitrile and dried under Ar. Figure 1c shows the representative transport characteristics of a $MoS_2$ FET before and after surface functionalization. In the pristine state, the width-normalized drain current ($I_D$) is ~5.5 × $10^{-3}$ µA/µm at zero gate voltage ($V_{GS}$), and the $I_D$ *vs.* $V_{GS}$ shows typical n-type transport[26,27]. After surface functionalization with DMAP-OED, the $I_D$ increases by 3 orders of magnitude to 3.6 µA/µm at $V_{GS}$ = 0 V, and the threshold voltage ($V_{th}$) shifts to a more negative $V_{GS}$, indicating strong n-type doping. The average 2D sheet carrier densities ($n_{2D}$) of the pristine and functionalized $MoS_2$ were extracted from the transfer curves (Fig. 1d) (See Supplementary Fig. S3 and Methods for details). The electron density increased from 3.5 ± 0.3 × $10^{11}$ $cm^{-2}$ to 1.9 ± 0.2 × $10^{13}$ $cm^{-2}$ ($\Delta n_{2D} = 1.88 \times 10^{13}$ $cm^{-2}$) on average as a result of doping from DMAP-OED. The high carrier density achieved in the functionalized $MoS_2$ corresponds to its degenerate limit where the Fermi level lies inside the conduction band[28].

As a control, we prepared a saturated solution of $BV^0$ in toluene and treated $MoS_2$ FETs for 10 minutes. The doping level achieved by the $BV^0$ functionalization is comparable to the reported values demonstrating that our method is consistent with other reported procedures[22]. As shown in Fig. 1d, the average electron density of $BV^0$-functionalized $MoS_2$ is 3 times lower than that of DMAP-OED-functionalized $MoS_2$. The lower doping power of $BV^0$ compared to DMAP-OED agrees with the relative redox potentials of the two molecules (Fig. 1a) as well as with their calculated doping powers. Additionally, control experiments show that the high n-type doping observed in the DMAP-OED-functionalized $MoS_2$ is not due to the acetonitrile solvent (Supplementary Fig. S4), nor to the formation of a continuous film of DMAP-OED that might contribute to electrical current (Supplementary Fig. S5).

Figure 2 shows the systematic increase in the carrier density of $MoS_2$ as the functionalization conditions were varied. Representative FET measurements for two different functionalization conditions are shown in Fig. 2a, which demonstrate that when $MoS_2$ is exposed to a DMAP-OED solution of higher concentration for a longer time there is a greater increase in $I_D$, which indicates stronger electron doping. Figure 2b provides a summary of FET measurements using various functionalization conditions, plotted in order of increasing average carrier density. As expected, increasing the treatment time or the concentration of DMAP-OED in solution increases the electron density of $MoS_2$; this can be attributed to more DMAP-OED molecules being present on the surface of $MoS_2$, which leads to higher n-type doping. In the case of a 10 mM DMAP-OED solution and an exposure time of 72 hours, the carrier density saturates at $\sim 5.8 \pm 1.9 \times 10^{13}$ cm$^{-2}$, representing the highest doping level achieved among all of the reported values of molecular doping in $MoS_2$ using organic or organometallic molecules[14-18,20,22].

The electron transfer from DMAP-OED to $MoS_2$ was further examined using XPS. Figure 3a and 3b show the Mo *3d* and S *2p* peaks of the pristine and functionalized $MoS_2$ using a 10 mM DMAP-OED solution and an exposure time of 10 minutes. After surface functionalization, the binding energies of the Mo and S core levels decrease by $\sim$ 0.6 eV and 0.5 eV, respectively, which indicates electron transfer from DMAP-OED to $MoS_2$, resulting in lower oxidation states of Mo and S[29]. The line shape of the XPS peaks did not change after functionalization, suggesting physisorption of DMAP-OED molecules on the surface of the $MoS_2$. Similar downshifts of the core level peaks were observed for functionalization of $MoS_2$ with $BV^0$ (Supplementary Fig. S6). Systematic XPS characterizations as a function of functionalization conditions are tabulated in Table S1. We also performed Raman spectroscopy and photoluminescence (PL) measurements on MoS$_2$ before and after DMAP-OED functionalization (Supplementary Fig. S7). Structurally MoS$_2$ remains in the semiconducting 2H phase while the PL intensity decreases significantly after functionalization in accordance with electron doping effects on MoS$_2$[30].

To measure the doping power of each DMAP-OED molecule to MoS$_2$, the measured increase in the carrier density must be divided by the number of dopant molecules on the MoS$_2$ surface. We quantitatively measured the number of DMAP-OED molecules on MoS$_2$ by examining the XPS N *1s* peak as the dopant molecule contains 4 nitrogen atoms (Fig. 3c). The N *1s* peak is absent in the pristine MoS$_2$, confirming its origin from DMAP-OED. By comparing the relative areas of the N and Mo peaks, a ratio of 1 DMAP-OED per 3.85 Mo atoms is obtained for MoS$_2$ functionalized with a 10 mM DMAP-OED solution for 10 minutes. Assuming a lattice constant of 3.16 Å for 2H-MoS$_2$[31], a surface density of $3 \times 10^{14}$ molecule/cm$^2$ was determined for DMAP-OED on MoS$_2$. For functionalization with BV$^0$, ~3 times as many BV$^0$ molecules were estimated to be on the surface of MoS$_2$ with a density of $9.6 \times 10^{14}$ molecule/cm$^2$ under similar functionalization conditions (Table S1). This is consistent with the smaller size of BV$^0$.

Given that the increase in the carrier density was $1.88 \pm 0.2 \times 10^{13}$ cm$^{-2}$ for the 10 mM DMAP-OED solution with 10 minute treatment time, the XPS analysis suggests that the doping power of DMAP-OED is ~ 0.06 electron per molecule, which is far lower than the prediction from DFT calculations as well as what is expected based on the redox potential of DMAP-OED. The lateral dimensions of a DMAP-OED molecule are about 0.97 nm × 1.08 nm, which is ~ 12 times larger than the unit cell of MoS$_2$. Therefore, a DMAP-OED molecule with its aromatic rings aligned parallel to the surface of MoS$_2$ would cover 12 Mo atoms, which suggests that the estimated ratio of 1 DMAP-OED per 3.85 Mo atoms from XPS indicates formation of multi-layer islands or films where molecules are stacked on top of each other. Since the doping power of the molecule is expected to decrease as it moves away from the surface of MoS$_2$, we estimate that not all the molecules measured by XPS are donating electrons equally to MoS$_2$.

The results described above suggest that when MoS$_2$ is functionalized with a 10 mM DMAP-OED solution for 10 minutes, a 'saturated' system in which many of the molecules are not donating electrons to MoS$_2$ is obtained. Thus, we examined the functionalization condition of a 0.1 mM DMAP-OED solution for 10 minutes. In this case, we could not use XPS to quantitatively measure the surface density of DMAP-OED on MoS$_2$ as the N *1s* peak was not clearly resolved after surface functionalization (Supplementary Fig. S6). We thus used AFM to estimate the molecule surface coverage and determine the doping power of DMAP-OED. Figures 3d and 3e show the AFM images of pristine and functionalized MoS$_2$ treated with a 0.1 mM DMAP-OED solution for 10 minutes, which show that the molecules aggregate and form islands. Analysis of AFM images shows an average areal coverage of 11% for functionalized MoS$_2$. The surface roughness of the uncovered areas is comparable to that of the pristine MoS$_2$, indicating that only the islands contain the molecules. Over 200 islands were analyzed to obtain the diameter and height distributions of the islands (Fig. 3f). A striking number of the islands are either ~ 0.8 nm or 1.3 nm in height, while they are broadly distributed in diameter with an average value of 21.4 ± 0.7 nm. Accounting for the height of the DMAP-OED molecule (0.375 nm, Fig. 1b), the space between the molecule and MoS$_2$ (0.209 nm), and the interlayer spacing between the molecules, the observed heights of 0.8 nm and 1.3 nm strongly suggest that the islands are either mono- or bi-layer islands. The surface coverage of DMAP-OED decreases to ~1.6 % with mainly monolayer islands when the functionalization time was shortened to 1 minute (Supplementary Fig. S8). We note that MoS$_2$ treated only with acetonitrile showed less than ~0.2 % areal coverage of small islands, eliminating the possibility that the islands are from the solvent molecules.

To calculate the doping power of a DMAP-OED molecule from the AFM data that shows mono- and bi-layer islands, we consider two scenarios: the first is electron doping by only the first layer of molecules closest to MoS$_2$ and the second is electron doping by both layers of molecules with equal doping power. We focus on the case of a 0.1 mM DMAP-OED solution functionalizing MoS$_2$ for 10 minutes. Based on the AFM results (Fig. 3d-f), the total number of DMAP-OED molecules in the first and second layer is estimated to be 1.25 × 10$^{13}$ molecules / cm$^2$ (Supplementary Information). Given the change in the carrier density of 7.9 ± 1.0 × 10$^{12}$ cm$^{-2}$ (Fig. 2b), the estimated doping power of a single DMAP-OED molecule to MoS$_2$ is 1.26 ± 0.15 electrons per molecule in the first scenario and 0.63 ± 0.08 electron per molecule in the second scenario. The doping power of DMAP-OED is on a similar order when the functionalization time is shortened to 1 minute, given the 10 times reduction in both the surface coverage and induced carrier density. Thus we observe a near unity doping efficiency for DMAP-OED to $MoS_2$.

DFT calculations were performed to provide further understanding of our results. Figure 4 shows the calculated electronic structure of DMAP-OED-functionalized $MoS_2$ for two cases: a single DMAP-OED molecule in an 8 × 8 $MoS_2$ supercell (1-layer (1L); Fig. 4a, c, e) and two molecules vertically stacked on $MoS_2$ in the same supercell (2-layer (2L); Fig. 4b, d, f). Electrons from the molecules are transferred predominantly to the interfacial region and Mo sites (Fig. 4a-b, Supplementary Fig. S9). These electrons are donated from the highest occupied molecular orbital (HOMO) of DMAP-OED. In the 1L case, the HOMO becomes approximately half-unoccupied and splits into two non-degenerate spin-polarized orbitals (Fig. 4e). The calculated doping power is 0.99 electrons per molecule to $MoS_2$ (Methods). In the 2L case, the HOMOs of the two molecules hybridize and split into two peaks in the projected density of states (PDOS) (Fig. 4f). One of the peaks is primarily localized on the top molecule and remains fully occupied after functionalization, while the other peak is primarily localized on the bottom molecule and becomes partially unoccupied (Fig. 4f). The two molecules donate a total of 1.27 electrons to $MoS_2$, with the molecule in direct contact with $MoS_2$ donating 0.86 electrons. These calculated doping powers are in good agreement with the experimental results. Further, the calculations show that increasing the surface coverage of DMAP-OED molecules by arranging two molecules side by side on $MoS_2$ (Supplementary Fig. S1) reduces the doping power per molecule, while the total charge donated by the molecules to $MoS_2$ increases. Table 1 summarizes the doping powers of DMAP-OED molecules with four different configurations in relation to $MoS_2$, and shows diminishing doping power per molecule for higher coverage. We note that the Fermi levels of the DMAP-OED-functionalized $MoS_2$ are within the conduction band of $MoS_2$ (Fig. 4e-f), consistent with the degenerate electron-doping observed experimentally. In the presence of a sulfur vacancy in $MoS_2$ (Supplementary Figure S1), we calculate that the doping power of DMAP-OED increases to 1.27 electrons per molecule in the 1L case, suggesting that the nature of $MoS_2$ can significantly influence doping efficiencies of DMAP-OED.

In conclusion, the ability to dope $MoS_2$ beyond its degenerate limit *via* an organic super electron donor provides opportunities to access correlated electronic phases, such as superconductivity at the degenerate level, and provides an alternative to ionic gating. Using DMAP-OED, we have demonstrated a record molecular doping level in monolayer $MoS_2$. Further, we established that a single DMAP-OED molecule donates 0.63–1.26 electrons per molecule to $MoS_2$, one of the first evaluations of the doping power of an OED. We observe that DMAP-OED aggregates and forms islands on $MoS_2$, which eventually limits the doping efficiency. Altogether, we establish that multiple factors, such as reduction potential, size, and binding mode of the dopant molecule, as well as interactions between the molecules and 2D materials and among the molecules themselves play an important role when considering the design of strong molecular dopants. In fact, forming a uniform dopant layer whereby all molecules contribute equally to doping should result in the doping efficiencies exceeding the values reported here. Overall, our findings provide insight to guide the development of strongly doped 2D materials.

**Methods:**
**Synthesis and characterization of DMAP-OED and $BV^0$.**
DMAP-OED was synthesized according to a literature procedure[32]. The $^1H$ NMR spectrum was consistent with that previously reported and the solid was stored under an $N_2$ or Ar atmosphere.
$BV^0$ was synthesized according to the following procedure:
To a 500 mL Schlenk flask in a glovebox under an $N_2$ atmosphere was added benzyl viologen dibromide (3.3 g, 6.75 mmol), magnesium powder (0.51 g, 20.25 mmol) and acetonitrile (30 mL). The Schlenk flask was sealed and removed from the glovebox and it was allowed to stir at 60 °C for five days. Initially, the reaction was a yellow suspension. After 30 minutes, the solution began to turn deep blue. The blue color deepened and the yellow precipitate disappeared over days. After five days, a red solution and red precipitate were present. The reaction flask was allowed to cool to room temperature and volatiles were removed under vacuum. The crude solid was taken up in 250 mL of THF and filtered under $N_2$, then the volatiles were removed under vacuum. The remaining solid was washed with room temperature ethanol (3x20 mL) under $N_2$ and then dried overnight under vacuum to afford $BV^0$ as a dark red solid, which was stored under

an $N_2$ atmosphere (400 mg, 18%). $^1$H NMR (500 MHz, THF-$d_8$) δ 7.29-7.28 (m, 8H), 7.22-7.19 (m, 2H), 5.71 (b, 4H), 5.25 (b, 4H), 4.19 (b, 4H). For the $^1$H NMR spectrum of $BV^0$, see Supplementary Fig. S11.

Benzyl viologen dibromide was synthesized according to a literature procedure[33]. Magnesium powder was activated by stirring it in tetrahydrofuran at 40 °C with 0.025 equivalents of 1,2-dibromoethane for one hour under a stream of $N_2$ with a gas outlet because the reaction evolves ethylene. The solid was then collected *via* filtration and washed further with tetrahydrofuran under an $N_2$ atmosphere. The solid was dried under vacuum and stored under an $N_2$ atmosphere. Acetonitrile used for synthesis of $BV^0$ and $MoS_2$ functionalization was purchased from Honeywell (Cat. No. CS017-56) and used without further purification.

**CVD synthesis of MoS$_2$.**

Monolayer $MoS_2$ flakes were synthesized on $SiO_2$/Si substrates in a 1-inch quartz tube furnace. ~0.4 mg of $MoO_3$ powder was placed in a quartz crucible at the center of the furnace, and ~200 mg of sulfur powder was placed upstream in an alumina crucible liner, with 17 cm separating the precursors. A 285 nm $SiO_2$/Si substrate was treated with piranha solution (3:1 $H_2SO_4$ : $H_2O_2$) for at least 1 hour, and subsequently rinsed with deionized water and dried using a stream of $N_2$ gas. The substrate was then treated with a single drop (~4 μL) of 100 μM perylene-3,4,9,10 tetracarboxylic acid tetrapotassium salt (PTAS). After drying the PTAS-treated substrate on a hot plate in air, the substrate was placed face down on the $MoO_3$ crucible. The quartz tube was purged several times with ultra-high purity Ar gas to ensure no residual oxygen was present. The furnace temperature was ramped to 850 °C, and then kept at 850 °C for 15 min, while flowing Ar at ~10 sccm. After the reaction, the furnace was naturally cooled to 580 °C, and then the lid was opened to accelerate the cooling to room temperature. Raman spectroscopy, photoluminescence (PL), AFM, and time-of-flight secondary ion mass spectrometry (TOF-SIMS) were utilized to confirm that the $MoS_2$ is a monolayer with uniform distributions of S and Mo (see Supplementary Fig. S2).

**Device fabrication and characterization.**

To avoid gate leakage, the as-grown $MoS_2$ monolayers were transferred to a fresh $SiO_2$ (285 nm) / Si ($p^+$) substrate, which served as back-gate for field-effect devices. For the transfer of $MoS_2$,

950 PMMA A4 (MicroChem) was spin-coated on the growth $SiO_2$ substrates containing $MoS_2$ flakes and baked at 120 °C for 5 minutes on a hot plate. The PMMA / $MoS_2$ film was released from the $SiO_2$ substrate by floating the sample in 2M KOH at 65 °C for one hour. The film was then rinsed with deionized water several times and transferred to a fresh $SiO_2$ / Si substrate. After drying the sample on a hot plate at 40 °C for 40 minutes, the sample was kept in acetone overnight to remove the PMMA. Electron beam lithography was used to pattern the metal contacts followed by e-beam evaporation of Ti (10 nm) / Au (60 nm) and lift-off in acetone overnight. Electrical measurements were performed on these devices before and after molecular functionalization. The electrical characteristics of the devices were measured in air using a semiconductor device analyzer (Agilent Technologies B1500A).

**Carrier density calculation.**
From the FET transfer curves, the 2D sheet carrier density in $MoS_2$ was calculated[22] by $n_{2D} = (I_D L)/(qWV_{DS}\mu)$, where $I_D$ is the drain current at the zero gate voltage; $L$ and $W$ are the length and width of the channel, respectively; $q$ is the electron charge, and $\mu$ is the field-effect mobility at $V_{DS}$ = 1 V. The mobility was calculated as $\mu = ((\partial I_D/\partial V_{GS}) L/(V_{DS}C_{ox}W))$, where $(\partial I_D/\partial V_{GS})$ is the maximum transconductance (see Supplementary Fig. S3), and $C_{ox}$ is the gate capacitance of $1.2 \times 10^{-8}$ F/cm$^2$ for 285 nm thick $SiO_2$ estimated based on the parallel-plate model. Numerous $MoS_2$ devices were measured to ensure reproducibility of the observed transfer curves and to obtain the average $n_{2D}$ and the standard error, which represent the error bars (Fig. 1d and 2b).

**XPS, PL, and Raman characterization.**
XPS was performed on a PHI VersaProbe II Scanning XPS Microprobe with an Al $K_\alpha$ monochromatic X-ray source. In order to prevent exposure of samples to the ambient environment, a vacuum vessel was used to transfer the samples from a glovebox to the XPS instrument. A beam spot with the diameter of 20 μm was used to obtain XPS data only from the $MoS_2$ flakes, assisted by scanning X-ray induced secondary electron imaging. All of the XPS spectra were calibrated using a carbon 1s peak located at 284.50 eV. PL and Raman measurements were conducted on a Horiba LabRAM HR Evolution Raman system with a 532 nm laser.

**AFM characterization.**

AFM images were acquired with a Cypher ES Environmental AFM System (Asylum Research Oxford Instruments) in tapping mode using the FS-1500AuD (Asylum Research Oxford Instruments) cantilever at a scan rate between 4.88 and 7.81 Hz. All imaging was performed in ambient conditions. To determine the surface coverage of DMAP-OED on functionalized flakes, AFM images were acquired from several scanning areas and with at least 5 different scan sizes between 500 nm and 5 μm. Images were also acquired from pristine $MoS_2$ flakes to determine their root-mean-square (RMS) surface roughness at a scan size of 500 nm. The RMS surface roughness of the uncovered areas of functionalized flakes was equivalent to that of the pristine $MoS_2$ surface, which confirms the absence of any molecules in these areas. The images were processed using ImageJ to set a background threshold, perform particle analysis, and calculate the percent surface coverage. For each set of functionalization conditions, the same analysis was performed for 15 samples and the results were averaged.

**DFT calculations.**

Spin-polarized DFT calculations for all functionalized $MoS_2$ monolayers were performed using the SIESTA code[34], with the generalized gradient approximation[35] for the exchange-correlation functional. The semi-empirical DFT-D2 method of Grimme[36] was used to describe the van der Waals interactions. Norm-conserving Troullier-Martins pseudopotentials with partial core corrections were used. Double-zeta plus polarization basis sets (details in Supplementary Information) were chosen to reproduce the experimental work function of $MoS_2$ and the trends in gas phase energy levels were predicted using Gaussian 16 (see Supplementary Information). The Brillouin zone was sampled by a 3×3×1 k-point grid. A mesh cutoff energy of 300 Ry was used to obtain the electronic wavefunctions and charge densities. Atomic positions were relaxed until the forces were smaller than 0.02 eV/Å. For calculating the doping power of the organic reductants, we performed Bader charge analysis[37] using a fine FFT mesh of 240×240×960, which was sufficient to achieve convergence. A finer k-point grid of 12×12×1 and a Gaussian broadening of 0.08 eV were used to obtain the PDOS.

## Acknowledgements

M. Y. acknowledges support from the Army Office of Research (W911NF-18-1-0367) for the device fabrication and FET measurements of surface-functionalized MoS$_2$. Synthesis of MoS$_2$ flakes was supported by NSF CAREER 1749742. Y. Z. acknowledges the Link Foundation energy fellowship. S. N. R. acknowledges the Ford Foundation for a graduate student fellowship. H. W. and N. H. acknowledge support of a seed grant from the Center for Research on Interface Structures and Phenomena at Yale University. J. B. C. thanks the NSF for the graduate research fellowship. Material characterizations were carried out at shared facilities including Yale West Campus Materials Characterization Core, the Yale Institute for Nanoscience and Quantum Engineering, and the School of Engineering Cleanroom. S. Y. Q., J. W. and K. A. U. acknowledge funding from Grant MOE2016-T2-2-132 from the Ministry of Education, Singapore, and support from the Singapore National Research Foundation, Prime Minister's

## Author Contributions

M. Y., Y. Z., H. W., N. H., and J. J. C. conceived the project. M. Y. and S. N. R. synthesized monolayer $MoS_2$ flakes, and fabricated and measured $MoS_2$ FET devices under the guidance of J. J. C. N. H., D. J. C., J. B. C. synthesized the DMAP-OED and $BV^0$. M. Y., Y. Z. and H. W. functionalized $MoS_2$ with the OEDs and carried out Raman, PL, and XPS measurements. S. N. R. carried out AFM characterizations. J. W. and K. A. U. performed and analyzed DFT calculations under the supervision of S. Y. Q. J.V.P. contributed to the development of $MoS_2$ synthesis and fabrication methods. D. J. H. carried out TEM characterizations. S. Y. participated in technical discussions. M. Y. and J. J. C. wrote the paper, with input from all authors.

## Corresponding Authors

Correspondence should be addressed to Judy J. Cha.

## Competing Interests

The authors declare no competing interests.

## Additional Information

**Supplementary information** for this paper is available at

Figure S1-S11, Table S1 and S2, and references. Atomic structures of all DMAP-OED/$MoS_2$ configurations considered in this study; Structural characterizations of $MoS_2$; Electrical characteristics of $MoS_2$ FET; Effect of the acetonitrile solvent on transport characteristics of the $MoS_2$ FETs; Electrical characteristics of blank devices; XPS surface analysis of $BV^0$ and DMAP-OED-functionalized $MoS_2$; Summary of the Molecule:Mo ratio and shift in the binding energies of Mo 3d and S 2p peaks for different functionalization conditions; Raman spectroscopy and PL characterizations; AFM surface analysis for $MoS_2$ functionalized with 0.1 mM DMAP-OED solution for 1 minute; Details of DFT calculations; Additional electronic structure information; Details about the $^1H$ NMR spectrum of $BV^0$

**Figures:**

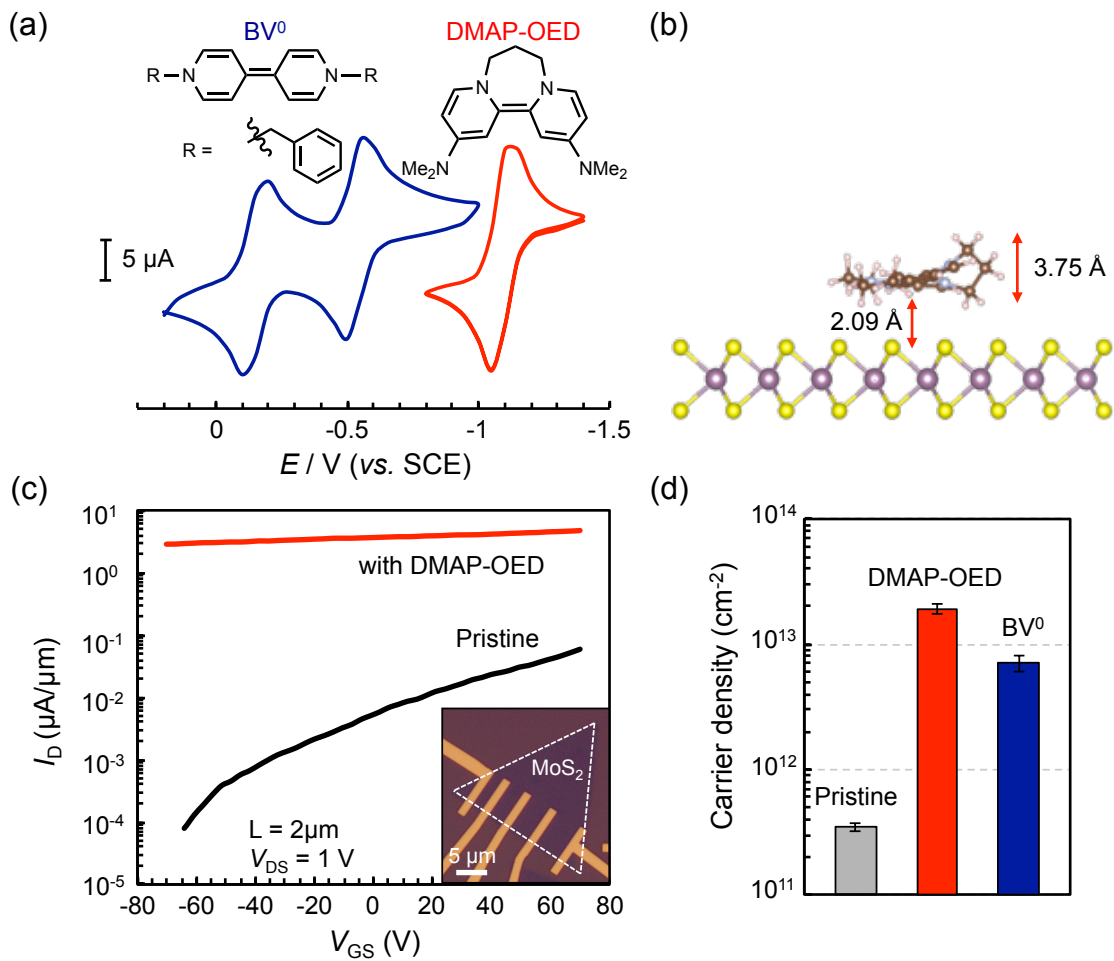

**Fig. 1 | Cyclic voltammograms (CVs) of the molecular dopants and transfer characteristics of MoS$_2$ transistors. a,** Chemical structures and CV scans of DMAP-OED and BV$^0$, indicating that DMAP-OED is a stronger reductant than BV$^0$. **b,** Atomic structure of a single DMAP-OED molecule on the surface of monolayer MoS$_2$. The aromatic rings of the DMAP-OED molecule are aligned parallel to the basal plane of MoS$_2$. **c,** Representative $I_D$–$V_{GS}$ transfer curves of a pristine MoS$_2$ FET and after functionalization with a 10 mM DMAP-OED solution for 10 minutes. The $I_D$ at $V_{GS}$ = 0 V increases by 3 orders of magnitude upon DMAP-OED doping and overall shows a weak gate dependency. The bias voltage is 1 V and the channel length is 2 μm. Inset: An optical image of the monolayer MoS$_2$ FET with various channel lengths. **d,** Average carrier densities at $V_{GS}$ = 0 V for pristine (gray), DMAP-OED (red), and BV$^0$ (dark blue) functionalized MoS$_2$ FETs prepared using a 10 mM molecular solution and 10 minute treatment time. The error bars in the carrier densities represent the standard error obtained from measuring 6 to 8 MoS$_2$ devices.

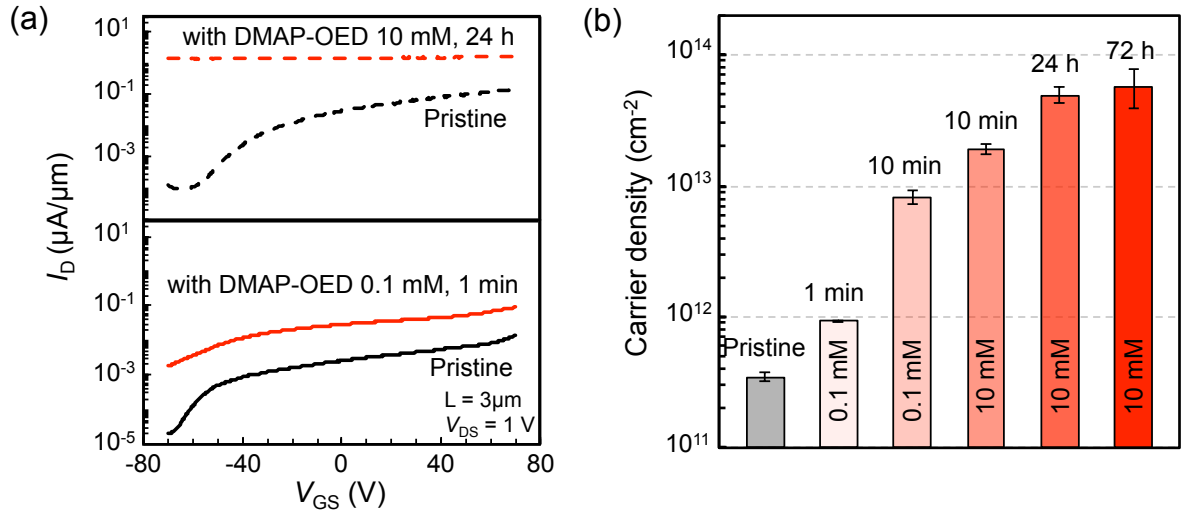

**Fig. 2 | Systematic functionalization of MoS$_2$ with DMAP-OED for increased electron doping. a,** Representative $I_D$−$V_{GS}$ transfer curves of MoS$_2$ FETs for two different DMAP-OED treatment conditions: (top) 10 mM, 24 hours and (bottom) 0.1 mM, 1 minute. The bias voltage is 1 V and the channel length of the devices is 3 μm. **b,** Systematic increase of the carrier density in MoS$_2$ at $V_{GS}$ = 0 V with increasing DMAP-OED concentration and functionalization time.

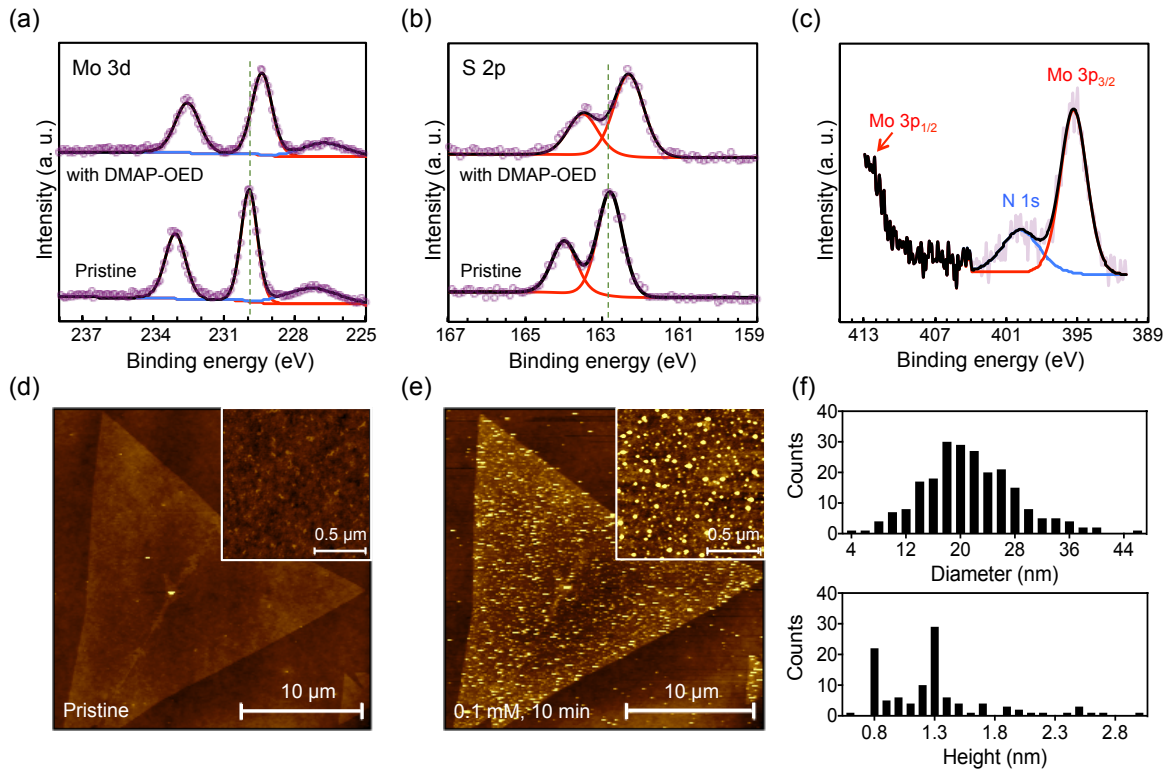

**Fig. 3 | XPS and AFM surface analysis of DMAP-OED-functionalized MoS$_2$. a-c,** XPS spectra of Mo *3d* **(a)**, S *2p* **(b)**, and N *1s* **(c)** core levels before and after treating MoS$_2$ with a 10 mM DMAP-OED solution for 10 minutes. The downshift of the binding energies of the pristine peaks, indicated by the green dashed lines in **(a)** and **(b)**, confirms lower oxidation states of Mo and S due to electron injection into MoS$_2$. **d, e,** AFM images of pristine **(d)** and DMAP-OED-functionalized MoS$_2$ after treating MoS$_2$ with a 0.1 mM DMAP-OED solution for 10 minutes **(e)**. Insets: Zoom-in AFM images. As revealed by the bright spots in **(e)** DMAP-OED tends to aggregate and form islands. **f,** Diameter and height distributions of the molecule islands formed on MoS$_2$ after functionalization with a 0.1 mM DMAP-OED solution for 10 minutes.

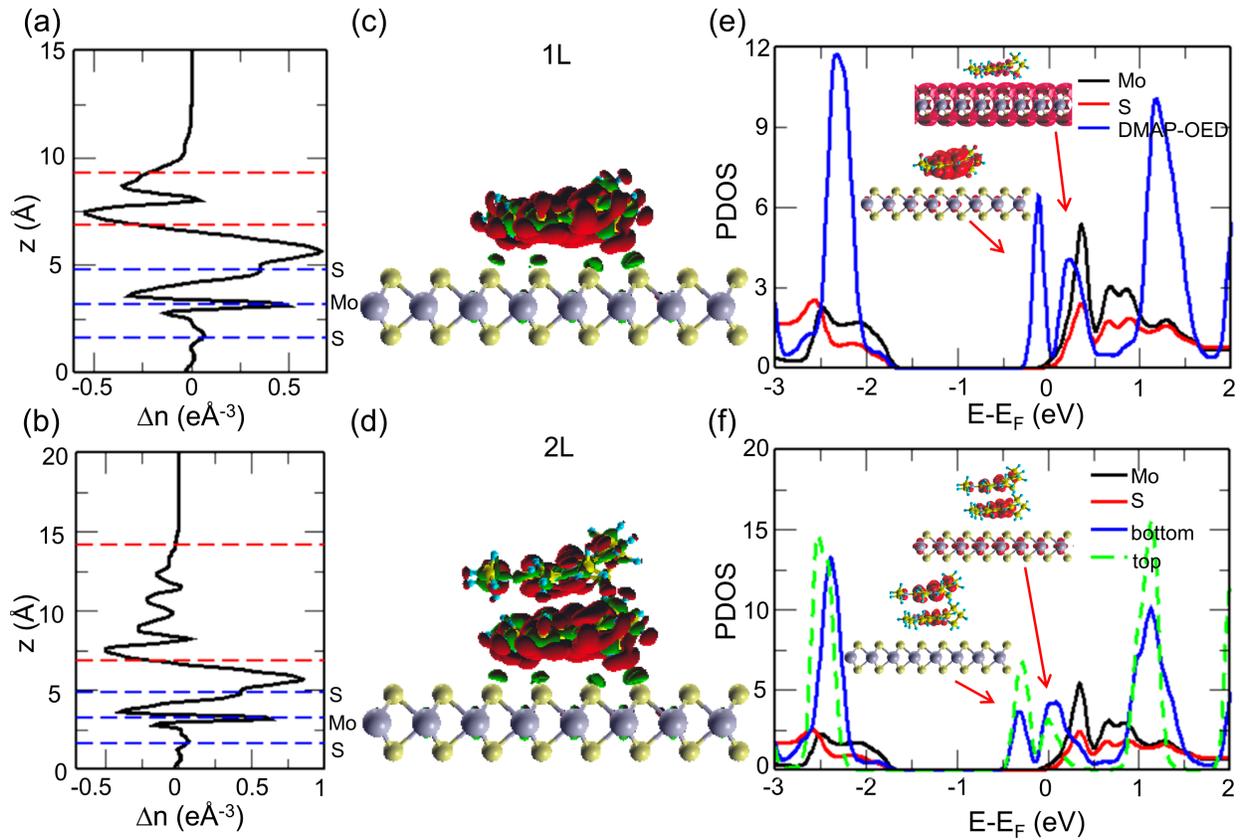

**Fig. 4 | Atomic and electronic structure of 1-layer (1L) and 2-layer (2L) configurations of DMAP-OED on monolayer $MoS_2$.** The planar-averaged differential charge density $\Delta n$ (e Å$^{-3}$) of **(a)** 1L and **(b)** 2L configurations. Blue dashed lines correspond to Mo and S atomic layers while red dashed lines denote the upper and lower boundaries of the molecules. Differential charge density $\Delta \rho$ of **(c)** 1L and **(d)** 2L configurations. The differential charge density is defined as $\Delta \rho = \rho_{mol/MoS_2} - \rho_{MoS_2} - \rho_{mol}$. Electron depletion and accumulation are indicated by red and green colors, respectively. The iso-value is 5% of the maximum value. Projected density of states (PDOS) for **(e)** 1L and **(f)** 2L configurations. The PDOS shown here is the sum of spin up and spin down PDOS (see Supplementary Figure S10 for spin-polarized PDOS). The PDOS on $MoS_2$ is divided by 64 for clarity. The partial charge density of selected peaks in the molecular PDOS are shown as insets (the iso-value is 1% of the maximum value in (e) and 5% of the maximum value in (f)).

**Table 1. Calculated number of electrons donated from each DMAP-OED molecule to monolayer MoS$_2$, in different atomic configurations.** The 1L configuration has one molecule in an 8 × 8 MoS$_2$ supercell (Fig. 4c), while the 2L configuration has two molecules stacked on top of one another in an 8 × 8 MoS$_2$ supercell (Fig. 4d). The '1L denser coverage' configuration has two molecules arranged side by side in an 8 x 8 MoS$_2$ supercell (Supplementary Fig. S9e), while the '2L denser coverage' configuration has four molecules in total in two layers where each layer has two molecules arranged side by side in an 8 x 8 MoS$_2$ supercell (Supplementary Fig. S9f). All atomic configurations are shown in Supplementary Fig. S1.

| Configurations | 1L (1 molecule) | 2L (2 molecules) | 1L denser coverage (2 molecules) | 2L denser coverage (4 molecules) |
|---|---|---|---|---|
| Molecule in 1$^{st}$ layer (in direct contact with MoS$_2$) | 0.99 | 0.86 | 0.89/0.89 | 0.74/0.74 |
| Molecule in 2$^{nd}$ layer | / | 0.41 | / | 0.24/0.24 |